\documentclass{elsarticle}
\usepackage{amsmath}
\usepackage{amssymb}
\usepackage{graphicx}
\newcommand{\be}{\begin{equation}}
\newcommand{\ee}{  \end{equation}}
\newcommand{\ba}{\begin{eqnarray}}
\newcommand{\ea}{  \end{eqnarray}}
\newcommand{\ve}{\varepsilon}

\begin{document}

\title{Correlation Widths in Quantum--Chaotic Scattering}

\author{B.~Dietz\fnref{ikp}}

\author{A.~Richter\corref{cor}\fnref{ikp,ect}}
\ead{richter@ikp.tu-darmstadt.de}

\author{H.~A.~Weidenm{\"u}ller\fnref{mpi}}

\address[ikp]{Institut f{\"u}r Kernphysik, Technische Universit{\"a}t
Darmstadt, D-64289 Darmstadt, Germany}
\address[mpi]{Max-Planck-Institut f{\"u}r Kernphysik, D-69029 Heidelberg,
Germany}
\address[ect]{$\rm ECT^*$, Villa Tambosi, I-38123 Villazzano (Trento), Italy}

\cortext[cor]{Corresponding author}

\date{\today}

\begin{abstract}

An important parameter to characterize the scattering matrix $S$ for
quantum--chaotic scattering is the width $\Gamma_{\rm corr}$ of the
$S$--matrix autocorrelation function. We show that the ``Weisskopf
estimate'' $d/(2\pi) \sum_c T_c$ (where $d$ is the mean resonance
spacing, $T_c$ with $0 \leq T_c \leq 1$ the ``transmission
coefficient'' in channel $c$ and where the sum runs over all channels)
provides a good approximation to $\Gamma_{\rm corr}$ even when
the number of channels is small. That same conclusion applies also to
the cross--section correlation function.

\end{abstract}

\maketitle

\section{Purpose}

Quantum--chaotic scattering is an ubiquitous phenomenon. It emerges
whenever Schr{\"o}dinger waves are scattered by a system with chaotic
intrinsic dynamics. Examples are the passage of electrons through
disordered mesoscopic samples, and compound--nucleus scattering.
Moreover, it occurs when
electromagnetic waves of sufficiently low frequency are transmitted
through a microwave cavity with the shape of a classically chaotic
billiard. In all these cases, chaotic scattering is due to the
numerous quasibound states of the system that appear as resonances in
the scattering process and that obey random--matrix statistics.

The generic approach to quantum--chaotic scattering~\cite{Mit10} is
based upon a random--matrix model for the resonances and, thus, for
the scattering matrix $S_{a b}(E)$, a function of energy $E$ where $a,
b$ denote the open channels. Within that approach, the energy
correlation function of the scattering matrix (the ensemble average
$\langle S^{}_{a b}(E - \ve/2) S^*_{c d}(E + \ve/2) \rangle$) can be
worked out analytically~\cite{Ver85} as a function of the energy
difference $\ve$, and approximate expressions for the cross--section
correlation function are also available~\cite{Dav88,Dav89,Die10}. The
correlation width of the cross section turns out to be rather close to
that of the scattering matrix in all cases~\cite{Die10}. That is why
we focus attention on the $S$--matrix correlation function in what
follows.

The analytical expression for the $S$-matrix correlation function is
given in terms of a threefold integral~\cite{Ver85}. The numerical
evaluation of that integral is quite cumbersome. To gain an
orientation of what to expect in a given situation, a simple
approximate expression for the width $\Gamma_{\rm corr}$ of the
$S$--matrix correlation function (and, by implication, of the cross
section) would therefore be helpful. A commonly used approximation
applicable in the regime of strongly overlapping resonances is the
``Weisskopf estimate''~\cite{Bla52}. It has for example been applied
to resonance spectra obtained from microwave experiments on quantum
chaotic scattering~\cite{Die10,Sch10}. The measurements were performed
in the regimes of isolated and weakly overlapping resonances and the
associated $S$ matrix comprised two dominant scattering channels and a
large number of weakly coupled ones. A motivation for the present
paper was to test the accuracy of the Weisskopf estimate under such
conditions. We will demonstrate that it provides a good approximation
for $\Gamma_{\rm corr}$ not only in the regime of strongly overlapping
resonances. For simplicity and without loss of generality we confine
ourselves to the case where the average $S$--matrix is diagonal,
$\langle S_{a b} \rangle = \langle S_{a a} \rangle \delta_{ ab}$. The
unitarity deficit of the average $S$--matrix is then measured by the
transmission coefficients $T_c = 1 - | \langle S_{c c} \rangle |^2$.
These obey $0 \leq T_c \leq 1$ for all $c$.

Naively, one might consider two alternatives for estimating
$\Gamma_{\rm corr}$. (i) The Weisskopf estimate expresses the total
average resonance width in terms of the mean resonance spacing $d$ 
and of the transmission coefficients $T_c$,
\be
\Gamma_{\rm W} = \frac{d}{2 \pi} \sum_c T_c \ . 
\label{1}
\ee
The sum in Eq.~(\ref{1}) runs over the open channels.

(ii) The ``Moldauer--Simonius sum rule''~\cite{Mol69,Sim74} gives the
following expression for the mean distance $(1/2) \langle \Gamma_\mu
\rangle$ of the poles of the scattering matrix (labeled by a running
index $\mu$) from the real energy axis.
\be
\langle \Gamma_\mu \rangle = - \frac{d}{2 \pi} \sum_c \ln [ 1 -
T_c ] \ .
\label{2}
\ee
For the case of unitary symmetry, the sum rule Eq.~(\ref{2}) has been
derived rigorously~\cite{Fyo99}. There is no reason to doubt that the
sum rule Eq.~(\ref{2}) holds also in the orthogonal case although a
proof exists only in fragmentary form~\cite{Som99}.

The width $\Gamma_{\rm W}$ in Eq.~(\ref{1}) and the double average
pole distance $\langle \Gamma_\mu \rangle$ as given by Eq.~(\ref{2})
agree whenever $T_c \ll 1$ for all $c$. In general, however, the
values of both quantities differ widely. For instance, for the case of
a single channel with $T \approx 1$, Eq.~(\ref{1}) yields $\Gamma_{\rm
W} \approx d/(2\pi)$ while Eq.~(\ref{2}) yields $\langle \Gamma_\mu
\rangle \gg d/(2\pi)$. An identification of $\Gamma_{\rm W}$ (of
$\langle \Gamma_\mu \rangle$) with the correlation width $\Gamma_{\rm
corr}$ would suggest that we deal with isolated (with strongly
overlapping) resonances, respectively. It is
known~\cite{Mol75,Bro81,Cel07} that Eq.~(\ref{2}) fails when any of
the $T_c$ is close to unity, and a comparison of the values of
$\Gamma_{\rm corr}$ given in the figures below with Eq.~(\ref{2})
confirms that fact. We ascribe that failure of the Moldauer--Simonius
sum rule to the fact that the fluctuation properties of the scattering
matrix depend not only on the location of the poles of $S$ but also on
the values of the residues. Little is actually known about the
latter~\cite{Fra00,Sch00}.

That leaves us with Eq.~(\ref{1}) as the only viable alternative. We
recall the conditions under which Eq.~(\ref{1}) is
obtained~\cite{Bla52}. One uses a time--dependent description and
considers a scattering system with constant resonance spacing $d_0$
coupled to a number of channels. The frequency with which a typical
wave function of the system approaches the entrance of a given channel
$c$ is $d_0 / h$, the probability with which the system escapes into
that channel is given by $T_c$, the partial width for decay into
channel $c$ is accordingly $d_0 T_c / (2 \pi)$. Summing over all
channels and postulating that the result applies also to systems that
do not have a constant resonance spacing $d_0$, one replaces $d_0$ by the
actual mean resonance spacing $d$ and arrives at Eq.~(\ref{1}). The
argument being semiclassical, one expects Eq.~(\ref{1}) to give an
approximate expression for the average resonance decay width in the
case of many channels or, more precisely, for $\sum_c T_c \gg 1$. 

That argument leaves open the question how $\Gamma_{\rm W}$ relates to
the correlation width $\Gamma_{\rm corr}$. In Ericson's
work~\cite{Eri} the identity of $\Gamma_{\rm W}$ and of $\Gamma_{\rm
corr}$ was postulated for $\sum_c T_c \gg 1$. A proof for that
assertion became available with the work of Ref.~\cite{Ver86}. There
it was shown that an expansion of the $S$--matrix correlation function
derived in Ref.~\cite{Ver85} in inverse powers of $\sum_c T_c$ yields
as the leading term a Lorentzian with width $\Gamma_{\rm W}$. This
result implies $\Gamma_{\rm corr} = \Gamma_{\rm W}$ in the Ericson
regime of strongly overlapping resonances, i.e., for $\sum_c T_c \gg
1$. A second case is that of many open channels each coupled weakly to
the resonances. Using the analytical expression for the $S$-matrix
correlation function~\cite{Ver85} Harney {\it et al.}~\cite{Harney}
showed that in that case, $\Gamma_{\rm W}$ also provides a good
approximation for $\Gamma_{\rm corr}$. Apart from these results for
the regimes of strongly overlapping and of isolated resonances coupled
to many channels no simple analytical expression exists for
$\Gamma_{\rm corr}$. In the present paper we investigate how much
$\Gamma_{\rm W}$ and $\Gamma_{\rm corr}$ differ for general values of
the number of channels and of the coupling strength in each
channel. We use the analytical result of Ref.~\cite{Ver85} for the
$S$-matrix correlation function to compute the width $\Gamma_{\rm
corr}$ numerically and compare the result with $\Gamma_{\rm W}$.

\section{Approach}

Starting point is the expression (see the review~\cite{Mit10})
\be
S_{a b}(E) = \delta_{a b} - 2 i \pi \sum_{\mu \nu} W_{a \mu}
[D^{-1}(E)]_{\mu \nu} W_{\nu b}
\label{3}
\ee
for the element $S_{a b}(E)$ of the scattering matrix connecting
channels $a$ and $b$, with
\be
D_{\mu \nu}(E) = E \delta_{\mu \nu} - H_{\mu \nu} + i \pi \sum_c
W_{\mu c} W_{c \nu} \ .
\label{4}
\ee Here $E$ is the energy. The real and symmetric matrix $H$ with
elements $H_{\mu \nu}$ and $\mu, \nu = 1, \ldots, N$ is a member of
the Gaussian orthogonal ensemble of random matrices (GOE). The
elements $H_{\mu \nu}$ are Gaussian--distributed random variables with
zero mean values and second moments given by $\langle H_{\mu \nu}
H_{\rho \sigma} \rangle = (\lambda^2 / N) [ \delta_{\mu \rho}
\delta_{\nu \sigma} + \delta_{\mu \sigma} \delta_{\nu \rho} ]$. The
matrix $H$ represents $N$ quasibound levels and their mutual
interaction. The parameter $\lambda$ has the dimension energy and
defines the average level spacing $d$ of the eigenvalues of $H$. In
the center of the GOE spectrum we have $d = \pi \lambda / N$. The
parameter $d$ defines the energy scale so that both $E$ and
$\Gamma_{\rm corr}$ are expressed in units of $d$. The real matrix
elements $W_{c \mu}$ couple the space of quasibound levels to
$\Lambda$ channels labelled $a, b, c, \ldots$. In the cases considered
in the present work the amplitudes for the passage from an intrinsic
state to a scattering channel coincide with those for the reverse
process, that is $W_{\nu c}=W_{c\nu}$.  Without loss of generality we
assume that $\sum_\mu W_{a\mu} W_{b\mu} = N v^2_a \delta_{a b}$. The
parameters $v^2_a$ define the mean strength of the coupling to channel
$a$. Since $H$ is random, the $S$--matrix is a matrix--valued random
process that depends on $E$. All moments and correlation functions of
$S(E)$ (defined by averaging over the GOE with the energy at or close
to the center of the GOE spectrum) depend only on the average
$S$--matrix elements $\langle S_{a b} \rangle$, on the transmission
coefficients $T_c$, and on energy differences. The latter are
expressed in units of $d$. With $x_a = \pi^2 v^2_a / d$ we have
\ba
\langle S_{a b} \rangle = \frac{1 - x_a}{1 + x_a} \delta_{a b} \ , \ \
T_a = \frac{4x_a}{(1 + x_a)^2} \ .
\label{5}
\ea
In Ref.~\cite{Ver85}, the autocorrelation and cross--correlation
functions of the elements of the $S$--matrix are given in terms of
these parameters. They are worked out for fixed $\Lambda$ in the limit
$N \to \infty$. We do not repeat the analytical expressions here.
These contain a threefold integration over real variables. We make use
of a simplification of these integrals in terms of variable
transformations first introduced in Ref.~\cite{Ver86} and summarized
in the Appendix of Ref.~\cite{Die10}. For a given set of transmission
coefficients $T_1, T_2, \ldots, T_\Lambda$ the resulting formula for
the $S$--matrix autocorrelation function
\be
C_{ab}(\epsilon) = \langle S^{}_{ab}(E - \ve/2)S_{ab}^*(E + \ve/2)
\rangle - \vert \langle S(E)_{ab}\rangle \vert^2 
\label{corr}
\ee
is evaluated numerically as a function of $\ve / d$. The full width at
half maximum of that function yields $\Gamma_{\rm corr} / d$.

\section{Results}

According to the Weisskopf estimate in Eq.~(\ref{1}), the correlation
width $\Gamma_{\rm corr}$ should depend only on the number of channels
$\Lambda$ and on the sum $T = \sum_c T_c$ of the transmission
coefficients. To test that assertion, we have for fixed values of
$\Lambda$ and of $T$ with $0 < T < \Lambda$ calculated the width of
the autocorrelation function Eq.~(\ref{corr}) for several sets of
parameters $\{T_1, T_2, \ldots, T_\Lambda \}$. These are subject to
the constraints $\sum_c T_c = T$ and $0 < T_c \leq 1$ and were
obtained with the help of a random--number generator. The number of
sets was typically between $25$ and $100$.  We evaluated
$C_{ab}(\epsilon)$, where $a$ and $b$ take either of the channel
numbers $1$ and $2$ and found that in the intermediate regime of
weakly overlapping resonances, the widths of $C_{12}(\epsilon)$, of
$C_{11}(\epsilon)$ and of $C_{22}(\epsilon)$ vary from set to set, in
contrast to the above assertion. The deviations of the ratios
$\Gamma_{\rm corr} / \Gamma_{\rm W}$ from unity are of comparable size
in all three cases. Therefore we show in the following only results
for $C_{12}(\epsilon)$.

To test the dependence of $\Gamma_{corr}$ on the values of $T_1, T_2$
associated with the incident and outgoing channels in the expression
for $C_{12}(\epsilon)$ we considered three cases. In case I, we chose
$T_1, T_2$ arbitrarily, that is, we did not sort the transmission
coefficients $\{T_1, T_2, \ldots, T_\Lambda \}$ by size. In case II
(case III) we ordered the transmission coefficients such that $T_1,
T_2$ take the maximal values (the minimal values, respectively) of all
$T$'s. In case II the channels $1$ and $2$ are the dominant ones, in
case III they are the most weakly coupled ones. Case II is relevant
for the microwave experiments described in
Refs.~\cite{Die10,Sch10,Die08}.

\begin{figure}[ht]
	\centering \includegraphics[width=8.5cm]{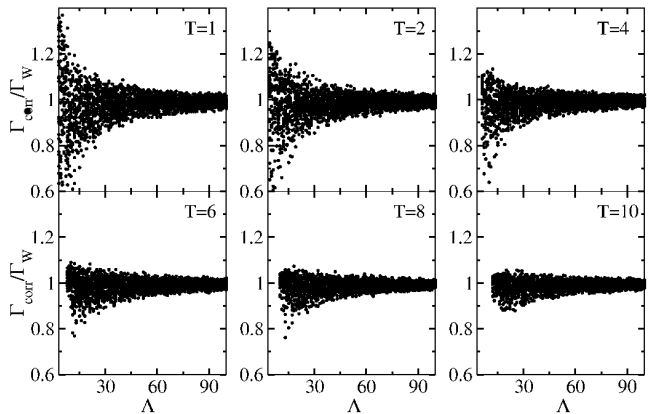}
	\caption{The ratio $\Gamma_{\rm corr} / \Gamma_{\rm W}$ (dots)
	versus the number $\Lambda$ of channels for several values of
	$T = \sum_c T_c$ as indicated in the panels. For each value of
	$\Lambda$ and of $T$, $25$ sets of the parameters $T_1, T_2,
	\ldots, T_\Lambda$ were randomly chosen. Each such set
	corresponds to a dot in the plot.}
\label{fig:BD1}
\end{figure}

In Fig.~\ref{fig:BD1} we consider case I and plot for several values
of $T$ given in the panels the ratio $\Gamma_{\rm corr} /
\Gamma_{\rm W}$ versus $\Lambda$ for the correlation function 
$C_{12}(\epsilon)$. The number of sets of $T_1, T_2, \ldots,
T_\Lambda$ chosen was $25$. To test the statistical significance of
the results we have increased the set size to $100$ and did not
observe a noticeable change. Each set corresponds to a dot in the
plot. The dots scatter about a mean value that is close to unity. For
fixed $T$ (fixed $\Lambda$), the width of the cloud of dots decreases
with increasing $\Lambda$ (increasing $T$, respectively). The width
indicates that in contrast to the Weisskopf formula, $\Gamma_{\rm
corr}$ does depend on the values of the individual transmission
coefficients. To further test this dependence we considered cases II
and III.  The ratios $\Gamma_{\rm corr} / \Gamma_{\rm W}$ resulting
from each of the 25 sets of transmission coefficients form clouds that
for both cases are very narrow as compared to those shown in
Fig.~\ref{fig:BD1}.  We do not display these as they would cover the
upper parts of the clouds shown in Fig.~\ref{fig:BD1} in case II, the
lower parts in case III.

\begin{figure}[ht]
	\centering \includegraphics[width=8.5cm]{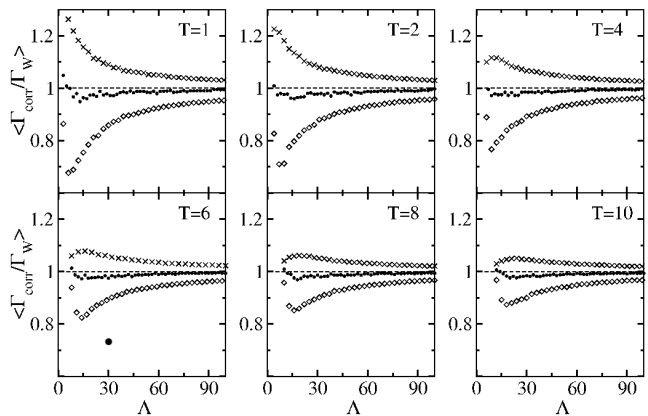}
	\caption{Averages of the ratio $\Gamma_{\rm corr} / \Gamma_{\rm
	W}$ over $25$ sets of transmission coefficients versus the
	number $\Lambda$ of channels for several values of $T$ as
	indicated in the panels. The center curves (dots) are for case
	I of the text (non-sorted transmission coefficients); the
	upper curves (crosses) for case II ($T_1, T_2$ maximal) and
	the lower curves (diamonds) for case III ($T_1, T_2$ minimal).}
\label{fig:BD2}
\end{figure}

\begin{figure}[ht]
	\centering \includegraphics[width=8.5cm]{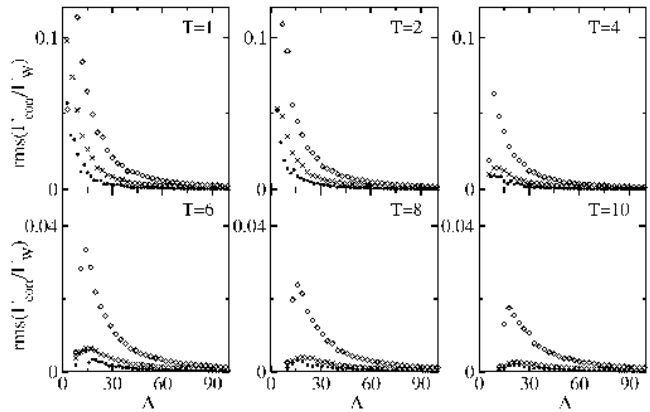}
	\caption{Same as Fig.~\ref{fig:BD2} but for the rms deviation
	from unity.}
\label{fig:BD3}
\end{figure}

Figures~\ref{fig:BD2} and \ref{fig:BD3} serve to quantify these
statements. In a plot similar to that of Fig.~\ref{fig:BD1},
Fig.~\ref{fig:BD2} shows the average of the ratios $\Gamma_{\rm corr}
/ \Gamma_{\rm W}$ over the $25$ realizations versus $\Lambda$ for case
I (center curve, circles), together with those for case II (upper
curve, crosses) and case III (lower curve, diamonds). For all values
of $\Lambda$ and $T$ the deviations of the average ratios from unity
are largest for case III and smallest for case I. In the latter case
the average ratio takes values above unity for small $\Lambda$ and
tends to values close to but below unity even for large $\Lambda$. In
contrast, for case II the average ratio is larger, for case III it is
smaller than unity for all values of $\Lambda$. In all three cases,
the deviations are largest for $\Lambda$ values between $10$ and $20$.
However, deviations from unity by more than $10$ percent occur only
for $T \lesssim 8$ or so. The curves rapidly tend to unity when
$\Lambda$ approaches $T$. Then all transmission coefficients take
values close to unity.

Figure~\ref{fig:BD3} shows similarly the root mean square (rms)
deviation of the ratio $\Gamma_{\rm corr} / \Gamma_{\rm W}$ from unity
(more precisely: the square root of the mean square deviation of the
ratios from unity). That quantity takes the largest values for case
III, and is very small unless $T \lesssim 8$. When $\Lambda$
approaches $T$ the rms values decrease rapidly for all three cases.

\begin{figure}[ht]
	\centering \includegraphics[width=8.5cm]{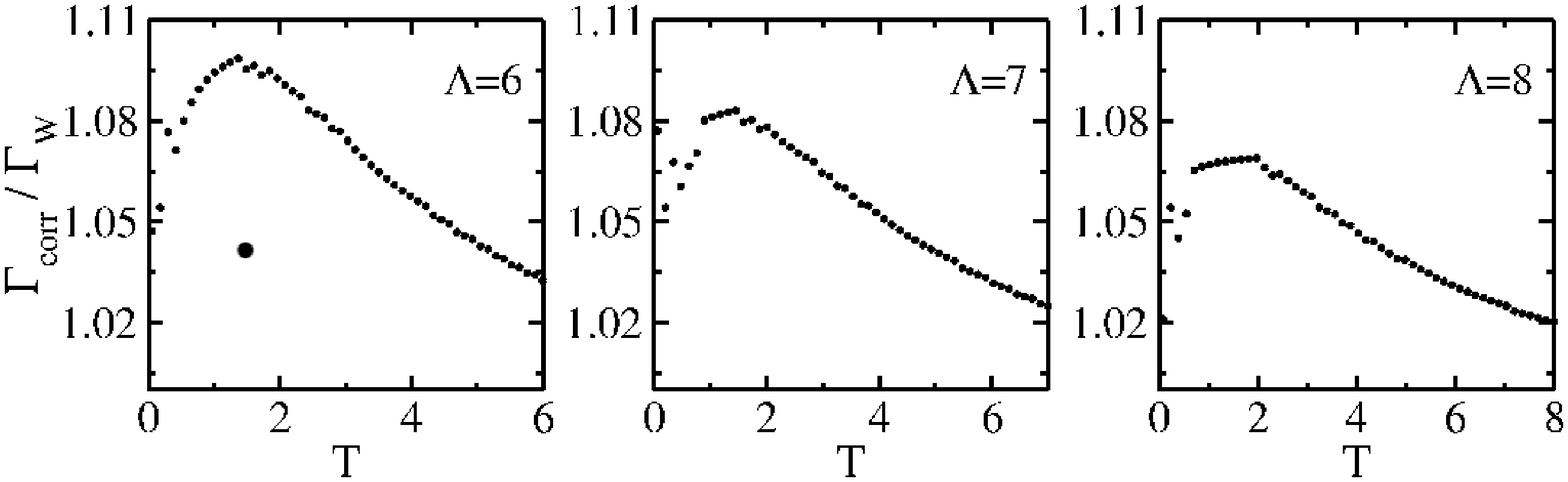}
	\caption{Ratio $\Gamma_{\rm corr} / \Gamma_{\rm
	W}$ (dots) for $\Lambda$ equal transmission coefficients versus
	$T$ for several numbers $\Lambda$ of channels as
	indicated in the panels.}
\label{figneu}
\end{figure}

\begin{figure}[ht]
	\centering \includegraphics[width=8.5cm]{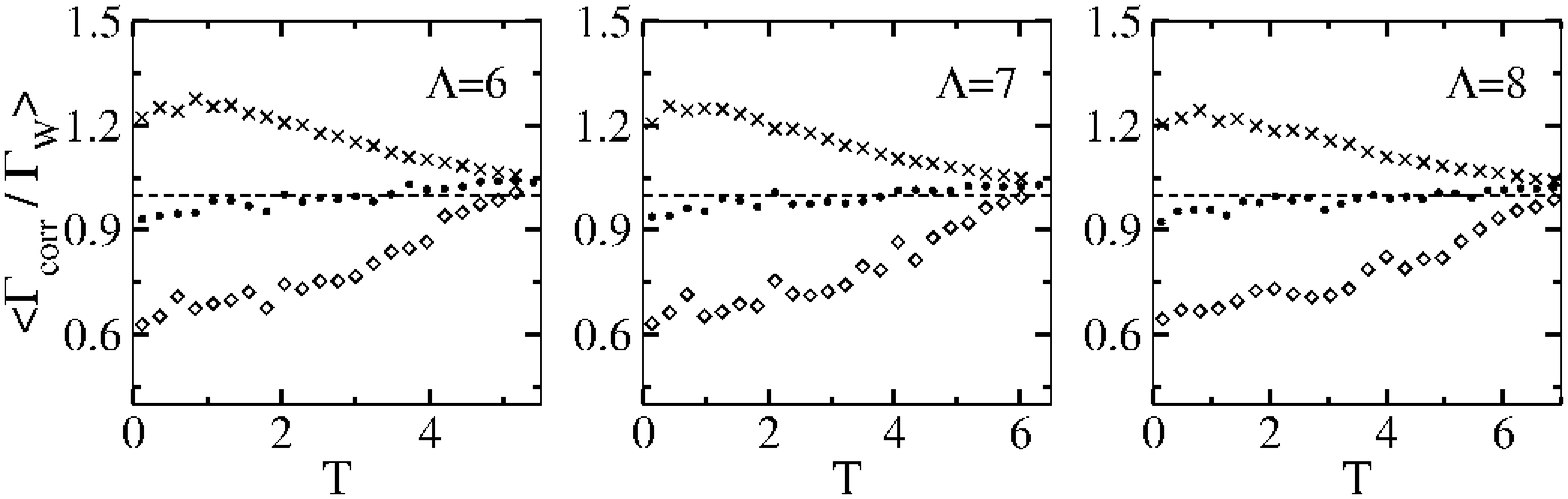}
	\caption{Average of the ratio $\Gamma_{\rm corr} / \Gamma_{\rm
	W}$ over $25$ sets of transmission coefficients versus $T$ for
	several numbers $\Lambda$ of channels as indicated in the
	panels and for case I (center curves, dots), for case II
	(upper curves, crosses) and for case III (lower curves,
	diamonds).}
\label{fig:BD4}
\end{figure}

\begin{figure}[ht]
	\centering \includegraphics[width=8.5cm]{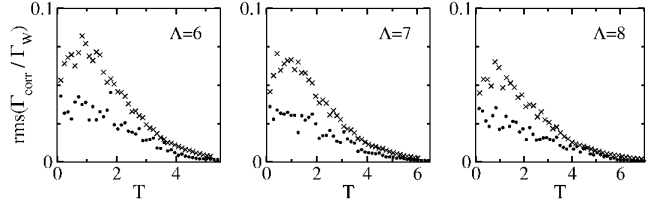}
	\caption{Root mean square deviations of the ratios
	$\Gamma_{\rm corr} / \Gamma_{\rm W}$ from unity versus $T$ for
	several numbers $\Lambda$ of channels as indicated in the
	panels for case I (center curves, dots) and for case II (upper
	curves, crosses). Case III is not shown.}
\label{fig:BD5}
\end{figure}

Because of the constraint $T\leq\Lambda$ only few points are at our
disposal in the regime of largest deviations. For a more thorough test
of the Weisskopf estimate we, therefore, also considered the case
where $\Lambda\leq 8$ is fixed and $T$ is varied. In Fig.~\ref{figneu}
we show the averages of the ratios $\Gamma_{\rm corr} / \Gamma_{\rm
W}$, with all transmission coefficients chosen equal, while in
Figs.~\ref{fig:BD4} and \ref{fig:BD5} cases I - III were considered as
done above in Fig.~\ref{fig:BD2}. Again the deviations of $\Gamma_{\rm
corr}/\Gamma_{\rm W}$ from unity are largest for case III and smallest
for case I. The average ratios are smaller than unity for all values
of $T$ for case III (lower curve, diamonds) while for case I they are
slightly smaller than unity for small $T$ and eventually reach a value
slightly above unity when $T$ approaches $\Lambda$, as is also
observed in Fig.~\ref{fig:BD2} for comparable values of $\Lambda$ and
$T$. For case II (upper curve, crosses) the deviations from unity are
less than $10$ percent unless $T\lesssim 3$; the ratio $\Gamma_{\rm
corr}/\Gamma_{\rm W}$ is above unity for all $T$. The dependence of
the curves on $\Lambda$ is similar to that for equally chosen
transmission coefficients in Fig.~\ref{figneu} although for given
values of $\Lambda$ and $T$ the latter deviate from unity much less
than the former. As in Fig.~\ref{fig:BD2} all curves in
Fig.~\ref{fig:BD4} tend to unity for $T$ close to $\Lambda$.

In Fig.~\ref{fig:BD5} we show the rms values for cases I and II. These
are less than $0.1$ for all values of $T$ and $\Lambda$ that were
considered. As suggested by Fig.~\ref{fig:BD4} the rms values for case
III are always larger than those for cases I and II but remain
smaller than $0.1$ for $T \gtrsim 3$. We do not show case III in
order not to blur Fig.~\ref{fig:BD4}.

\begin{figure}[ht]
	\centering \includegraphics[width=8.5cm]{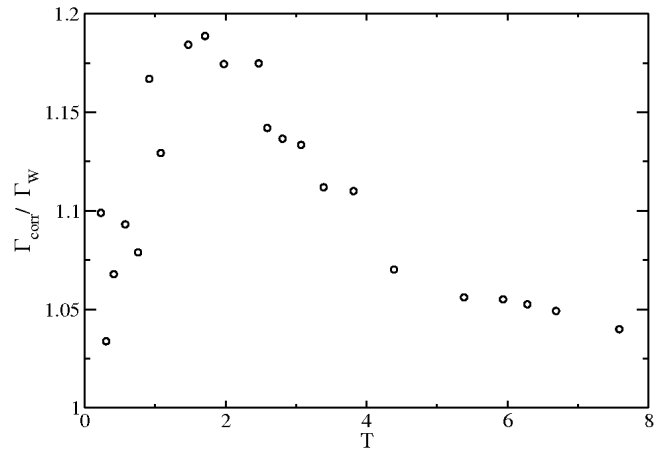}
	\caption{Ratio $\Gamma_{\rm corr} / \Gamma_{\rm
        W}$ for several sets of transmission coefficients taken from a microwave
        experiment~\cite{Die10,Die08} versus $T$. The sets consist of two dominant channels $1$ 
        and $2$ and $300$ weakly coupled ones, i.e., $T_c\ll T_1, T_2$ for
        $c\geq 3$.}
\label{fig:BD6}
\end{figure}

We also compared the Weisskopf estimate with data from a microwave
experiment described in Refs.~\cite{Die10,Die08}. The transmission
amplitudes $S_{12}, S_{21}$ and the reflection amplitudes $S_{11},
S_{22}$ of microwaves coupled into and out of a flat resonator via
two antennas $1, 2$ were measured. The resonator had the shape of a
tilted stadium billiard whose classical dynamics is chaotic
\cite{tilted}. The transmission coefficients $T_1\simeq T_2$
associated with the two antennas were determined via the relation
$T_c=1-|\langle S_{c c}\rangle|^2$ from the reflection spectra. Here
the angular bracket denotes a frequency average. To this end the
spectra, which exhibit isolated resonances for low frequencies and
increasingly overlapping resonances with increasing frequency, were
devided into $22$ frequency intervals of equal size. Dissipation into
the walls of the resonator was accounted for by introducing $300$
weakly coupled fictitious channels with equal transmission
coefficients $T_c$ where $T_c \ll T_1, T_2$ for $c \geq 3$. Hence
channels $1$ and $2$ are the dominant ones in these experiments, as in
case II. The transmission coeffcients of the absorptive channels were
determined from a fit of the analytic expression for the
autocorrelation function $C_{12}(\epsilon)$ to the experimental one
\cite{Sch10}. The width of the best fit yields $\Gamma_{\rm corr}$
while $\Gamma_{\rm W}$ is computed from the resulting transmission
coefficients and Eq.~(\ref{1}). From the measured spectra we thus
obtained altogether $22$ values for the ratios $\Gamma_{\rm
corr}/\Gamma_{\rm W}$. These are shown in Fig.~\ref{fig:BD6}. For
$T\lesssim 1$ they are very close to unity, as expected for a large
number of weakly coupled channels. They approach a maximum for
$T\simeq 2$ and then decrease again. For $T\gtrsim 4$ the deviations
of the ratios from unity are around $5$ per cent. The ratios are
larger than unity for all values of $T$, in agreement with the
numerical results for case II. Thus, in these experiments $\Gamma_{\rm
W}$ underestimates the correlation width. The deviations of the
Weisskopf estimate $\Gamma_{\rm W}$ from $\Gamma_{\rm corr}$ are
similar for the sets of transmission coefficients resulting from other
microwave experiments described in Ref.~\cite{Sch03}. In another
experiment~\cite{Alt1995} a superconducting chaotic microwave billiard
was used, and dissipation by the walls was negligible. In that
experiment three attennas were attached to the resonator so that
$\Lambda = 3$. In Ref.~\cite{Alt1995} the transmission coefficients
are determined from the partial widths of the measured resonances.
Their sum $T\simeq 0.55$ yields $\Gamma_{\rm W}$ while $\Gamma_{\rm
corr}$ is determined from the experimental autocorrelation function
shown in Fig.~4 of Ref.~\cite{Alt1995}. This yields $\Gamma_{\rm corr}
/ \Gamma_{\rm W} \simeq 1.13$, in good agreement with our numerical
results for small values of $\Lambda$ and $T<\Lambda$.

From the results obtained with the non-sorted transmission
coefficients we conclude that the Weisskopf estimate Eq.~(\ref{1})
constitutes a good approximation to $\Gamma_{\rm corr}$ for
practically all values of $\sum_c T_c$. Maximal deviations occur for
small values of $\Lambda$ and $T$ unless $\Lambda\simeq T$. That
statement applies also when incident and outgoing channel are the
dominant ones (case II). The largest deviations are observed when
these are the most weakly coupled channels (case III). In cases I and
II relative deviations of $\Gamma_{\rm corr} / \Gamma_{\rm W}$ from
unity larger than $10$ percent are only observed for $T \lesssim
4$. Generally, for $T\simeq\Lambda$ or for $T\gtrsim 8$ the deviations
are less than $10$ percent and decrease rapidly with increasing $T$ or
$\Lambda$. We have shown that our results are relevant for the
microwave experiments described in
Refs.~\cite{Die10,Die08,Sch03,Alt1995}. It would be interesting to
perform similar tests on the data obtained, for instance, in the
experiments in Refs.~\cite{Anlage,Sirko}.

\section*{Acknowledgement} This work was supported through the SFB~634
by the DFG.

\end{document}